\documentclass[reprint,aps,prb,superscriptaddress,amsmath,amssymb,twocolumn,floatfix]{revtex4-2} 

\usepackage{listings}
\usepackage{footmisc}
\usepackage{braket}
\usepackage{graphicx}
\usepackage[caption=false]{subfig}
\usepackage[colorlinks=True,linkcolor=red,citecolor=blue,urlcolor=blue]{hyperref}

\usepackage{blkarray}
\usepackage{array}
\usepackage[dvipsnames]{xcolor}
\usepackage[normalem]{ulem}

\usepackage{xspace}

\usepackage{tabularx}
\usepackage{array}   
\newcolumntype{L}{>{$}l<{$}} 
\newcolumntype{R}{>{$}r<{$}} 
\newcolumntype{C}{>{$}c<{$}} 

\usepackage{dsfont}
\newcommand{\ii}{{\mathrm{i}}} 
\newcommand{\e}{{\mathrm{e}}} 
\usepackage{bm}
\renewcommand{\vec}[1]{\bm{#1}} 

\newcommand{\1}{\mathds{1}} 
\newcommand{\trans}{\mathrm{T}}

\definecolor{darkgreen}{rgb}{0.02,0.45,0.0}

\newcommand{\LCO}{La$_2$CuO$_4$}
\newcommand{\SCOCl}{Sr$_2$CuO$_2$Cl$_2$}
\newcommand{\SRCOCl}{SrRbCuO$_2$Cl$_2$}

\date{\today}

\begin{document}
    \title{Exploring d-Wave Magnetism in Cuprates from Oxygen Moments}
\author{Ying Li}
\thanks{These two authors contributed equally}
\affiliation{Physics Department, Technical University of Munich, TUM School of Natural Sciences, 85748 Garching, Germany}
\affiliation{MOE Key Laboratory for Nonequilibrium Synthesis and Modulation of Condensed Matter, School of Physics, Xi'an Jiaotong University, Xi'an 710049, China}
\author{Valentin Leeb}
\thanks{These two authors contributed equally}
\affiliation{Physics Department, Technical University of Munich, TUM School of Natural Sciences, 85748 Garching, Germany}
\affiliation{Munich Center for Quantum Science and Technology (MCQST), Schellingstr. 4, 80799 München, Germany}
\author{Krzysztof Wohlfeld}
\affiliation{Institute of Theoretical Physics, Faculty of Physics, University of Warsaw, Pasteura 5, PL-02093 Warsaw, Poland}
\author{Roser Valent{\'\i}}
\affiliation{Institut f\"ur Theoretische Physik, Goethe-Universit\"at Frankfurt,
Max-von-Laue-Strasse 1, 60438 Frankfurt am Main, Germany}
\author{Johannes Knolle}
\affiliation{Physics Department, Technical University of Munich, TUM School of Natural Sciences, 85748 Garching, Germany}
\affiliation{Munich Center for Quantum Science and Technology (MCQST), Schellingstr. 4, 80799 München, Germany}
\affiliation{Blackett Laboratory, Imperial College London, London SW7 2AZ, United Kingdom}

	\date{\today}
	
	\begin{abstract}
The antiferromagnetic parent phase of high-T$_c$ cuprates has been established as a N\'eel state of copper moments, but early work pointed out the important role of ligand oxygen orbitals. Using the three-orbital Emery model, we explore how, and under which conditions, doping-induced antiferromagnetic ordering of weak magnetic moments on the oxygen sites can lead to unconventional d-wave magnetism with spin-split electronic bands. The mechanism for forming such altermagnetic (AM) states in cuprates does not rely on a lowering of the crystal symmetry but rather on interaction-induced formation of magnetic moments on directional oxygen orbitals within the crystallographic unit cell. Therefore, we obtain two different types of AM, namely a (0,0)-AM and a ($\pi$,$\pi$)-AM. We explore different regimes and challenges for realizing oxygen AM supported by Hartree-Fock calculations and complementary exact diagonalization of small clusters. While the region of interacting parameters needed to realize these states may be difficult to achieve in known high-T$_c$ cuprates, we propose a scenario to realize AM induced by oxygen magnetic moments in a cuprate-based candidate compound using density functional theory and discuss experimental implications.
\end{abstract}
    
\maketitle
	
\section{Introduction}
Understanding the parent phase of cuprates has been key for elucidating the emergence of high-T$_c$ superconductivity~\cite{bednorz1986possible,anderson1987resonating}. The insulating parent N\'eel state is 
typically
described as a commensurate antiferromagnet (AFM) of a single orbital model of magnetic moments on the copper sites~\cite{Gros1994,Dagotto1994, Imada1998}. However, early on Emery discussed the relevance of oxygen orbitals for understanding the magnetic fluctuations upon doping which are responsible for pairing~\cite{emery1987theory}. 
Not only is the exchange between copper moments mediated via the directional $p_x$, $p_y$ orbitals~\cite{emery1987theory,Zhang1988,Jefferson1992,
Weber2014, Sheshadri2023} but the position of oxygen on the bond centers leads to a unit cell with an internal structure which allowed for the prediction of intra-unit cell order, for example of spin nematic~\cite{fischer2011mean,fischer2014nematic} or loop current type~\cite{varma2006theory,chakravarty2001hidden}. Neutron scattering experiments established the importance of the hybridization between oxygen and copper orbitals for understanding the magnetic form factors~\cite{walters2009effect}. Famously, in the pseudo-gap phase neutron scattering measurements also observed magnetism without additional lattice symmetry breaking pointing to intra-unit cell order~\cite{fauque2006magnetic,li2008unusual}, which was argued to be consistent with staggered magnetic moments of the oxygen sites. Motivated by these experiments, we aim to understand the mechanism for stabilizing magnetism on oxygens, as well as the resulting unique properties of the magnetic states, in the context of cuprates described by the basic Emery model. 
\begin{figure}
\center
\includegraphics[width=0.9\linewidth]{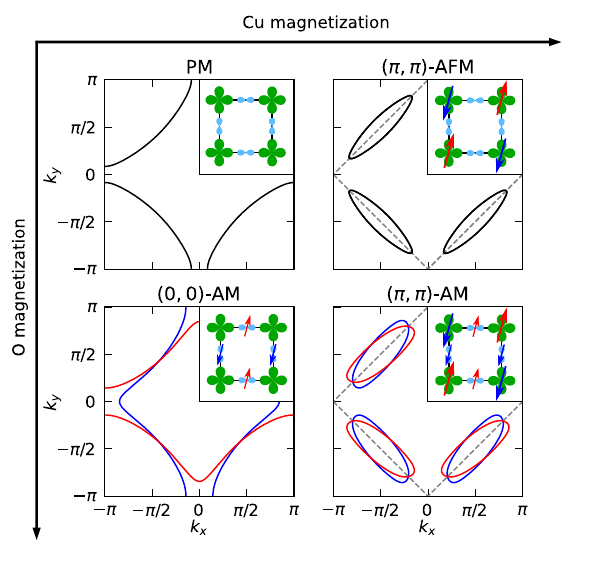}
\caption{Typically the AFM phase of cuprates feature AFM moments on the Cu atoms (x-axis). Here, we additionally explore the formation of AFM moments on the oxygen orbitals (y-axis), which leads to two different types of AM states. The insets of the subpanels show the possible magnetic phases we obtain (Cu green, O light blue). The subpanels show the spin resolved Fermi surface (FS) for the corresponding phase (black degenerate, red spin-up, blue spin-down). The magnetic Brillouin zone is shown as gray dashed square. Note that the lower left intra-unit cell ordering does not break translational symmetry as suggested in previous neutron scattering experiments~\cite{fauque2006magnetic,li2008unusual}.}
\label{fig1}
\end{figure}

Recently, a new type of collinear antiferromagnetism, so called altermagnetism (AM) based on a spin-symmetry classification has been proposed~\cite{Libor2020, Mazin2021, Libor2022,Libor2022a, Libor2022c} which is a unified way to describe previous reported anomalous behavior in magnetic systems~\cite{ahn2019antiferromagnetism,Hayami2019,Yuan2020, Ma2021}. AM combines  spin-split electronic bands expected for ferromagnets (FM) with a vanishing net moment of AFMs.
The reason is the absence of translational (or inversion) symmetry between the AFM sublattices, which can either originate from a low crystal symmetry or appear spontaneously from directional orbital ordering~\cite{leeb2024spontaneous}. In the context of AMs, the cuprate \LCO\ has been proposed as a candidate material~\cite{Mazin2022,Libor2022}  arising from a reduced crystal symmetry due to the canting of CuO$_6$ octahedra. However, the resulting spin splittings are small and disappear when the structure becomes tetragonal under doping. In three-dimensional analogues of the parent compound CuAg(SO$_4$)$_2$, AM appears, again, because of crystallographically different Cu-Ag sublattices~\cite{Jeschke2024}. 

Here, we show that AM of the $d$-wave type can already appear in highly symmetric tetragonal CuO$_2$ planes if magnetic moments can be stabilized on directional oxygen orbitals. 
Typically, cuprates are modeled to have AFM moments on the Cu atoms, see upper right panel of Fig.~\ref{fig1}. Here, we explore  the possibility of  additional AFM moments on the oxygen orbitals (lower panels), which can result in two distinct types of AM states: First, a (0,0)-AM which does not break any translational symmetry but leads to a spin splitting because of the intra-unit cell AFM arrangement. Second, a ($\pi$, $\pi$)-AM can appear where both Cu and O have AFM moments and the unit cell is spontaneously enlarged leading to a backfolding of the electronic bands (see right panels). The interesting point is that for both scenarios intra-unit cell oxygen AFMs is responsible for distinct spin-split electronic bands with a $d$-wave form factor as shown in the lower panels of Fig.~\ref{fig1}. 

We note that  AM without translational symmetry breaking has been discussed previously~\cite{fischer2011mean,Bose2024,giuli2024altermagnetism} and, indeed, the spin-nematic mean-field state identified in  Ref.~\cite{fischer2011mean} is our (0,0)-AM. Here, we concentrate on exploring the microscopic origin of oxygen-induced cuprate AM, the microscopic parameter regime and its unique electronic properties.
We identify three distinct microscopic mechanisms that could lead to the formation of magnetic moments on the oxygens: (i) A direct exchange between oxygen due to a large repulsive interaction $U_p$ on oxygen sites; (ii) A strong charge transfer bringing the $p$-oxygen orbitals closer to the Fermi energy; and (iii) A large oxygen-oxygen hopping $t_{pp}$ effectively shifting magnetic moments of copper to oxygens. 

An important question concerns the parameter regime in which oxygen moments appear. We find that for oxygen-oxygen hoppings and Coulomb repulsion values usual for cuprates the ligand sites remain non-magnetic. However, using Hartree-Fock mean-field theory (MFT) and complementary exact diagonalization (ED) studies of small clusters, we find that slightly larger values, which are potentially relevant for certain families of Cu-O-based materials, d-wave magnetism could be relevant. 

Our work is organized as follows. We explore the emergence and stability of distinct AM phases summarized in Fig.\ref{fig1} within the three-band Emery model, which we introduce in detail. We identify distinct mechanisms on how AFM correlations between oxygen moments can be stabilized,  which we support by MFT and ED of finite clusters. We then propose a potential cuprate material realizing AM using density functional theory (DFT). Finally, we discuss the unique experimental properties of the putative $d$-wave magnetism like anisotropic spin transport and splitting of spin wave branches observable via inelastic neutron scattering experiments. 

\section{Electronic structure and model Hamiltonian.}
All cuprates feature CuO$_2$ planes with Cu$^{2+}$ ions in a 3$d^9$ configuration with a localized hole in the Cu 3$d_{x^2-y^2}$ orbital. The parent compounds are charge-transfer insulators~\cite{Zaanen1985}, where the top of the band formed by the oxygen $p_x$, $p_y$ orbitals is located above the lower Hubbard band. This is well captured within the three-band Emery model~\cite{emery1987theory, Weber2014, Sheshadri2023}. 
Under doping, holes transfer to the oxygen ligands and 
at large doping
a metallic state emerges~\cite{Uchida1991} which is also consistent  with DFT calculations~\cite{Czyzyk1994,Pesant2011,Furness2018,Bielmann2002, Lane2018}. 

In this work we concentrate on the three-band Emery model
\begin{align}
\mathcal{H}=&\sum_{i\mu\sigma}\epsilon_{\mu}n_{i\mu\sigma}+\sum_{i\mu j\nu \sigma}t_{i\mu j\nu}\mathbf{c}_{i\mu\sigma}^\dagger \mathbf{c}_{j\nu\sigma} \nonumber\\
&+\sum_{i\mu} U_{\mu} n_{i\mu\uparrow}n_{i\mu\downarrow},
\label{eq:hamiltonian}
\end{align}
where $i, j$ label the square lattice unit cell, consisting of the $d_{x^2-y^2}$ Cu orbital ($\mu, \nu = d$) which is surrounded  by O $p_x$ and $p_y$ orbitals ($\mu, \nu = p_x, p_y$). $\mathbf{c}_{i\mu\sigma}^\dagger$ ($\mathbf{c}_{j\nu\sigma}$) create (annihilate) holes with orbital $\mu$ ($\nu$) and spin $\sigma$. $n_{i\mu\sigma}$ is the corresponding number operator. $U_{\mu}$ is the Hubbard type Coulomb repulsion for orbital $\mu=d,p$. $\epsilon_{\mu}$ is the orbital on-site energy, and $t_{i\mu j\nu}$ is the tunneling matrix element between $i\mu$ and $j\nu$. Fig.~\ref{fig:MFT_phases1} (a) shows the dominant hopping parameters in the CuO$_2$ plane, where $t_{pd}$ and $t_{pp}$ are the nearest neighbour Cu-O, and O-O hoppings, respectively. The parameters in the hole picture are obtained as $\epsilon_d$ = 0 eV, $\epsilon_p$ = 2.2 eV, $t_{pd}$ = 1.3 eV, and $t_{pp}$ = 0.6 eV for \LCO\  from projective Wannier functions~\cite{Eschrig2009} applied to non-relativistic full potential local orbital (FPLO) calculations~\cite{Koepernik1999}\footnote{We extracted the parameters  employing the Perdew-Burke-Ernzerhof generalized gradient approximation (GGA)~\cite{PBE1996} as exchange-correlation functional and a mesh of 9 $\times$ 9 $\times$ 5 {\bf k} points in the first Brillouin zone (FBZ).}. Considering the electronic structure of copper oxides, at charge neutrality the three orbital unit cell includes five electrons and one hole per unit cell such that the $d$-electron band is half filled. Hole doping is quantified by $\delta$, e.g. $4.5$ (4) electrons or $1.5$ (2) holes per unit cell means $\delta = 0.5$ ($1$). 

\section{AM driven by repulsion on oxygen $U_p$}
\begin{figure}
\includegraphics[width=\linewidth]{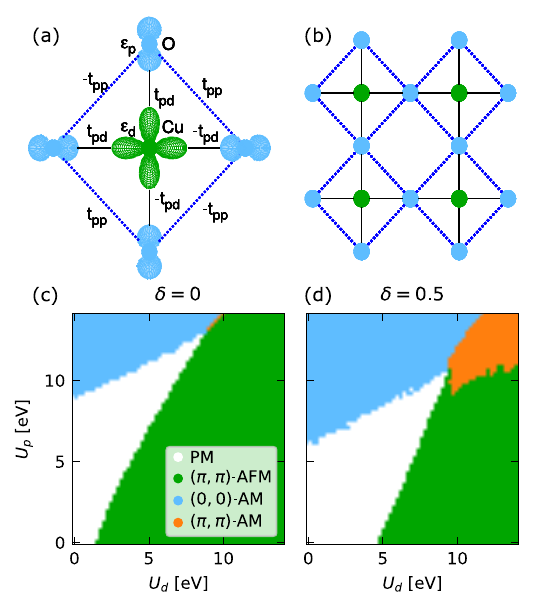}
\caption{Illustration of the hopping paths of the three-orbital Emery model (a), and the cluster considered in ED (b). MFT phase diagram for  undoped (c) and hole doped (d) case as a function of the Cu and O interaction strengths $U_d$ and $U_p$. Hole doping is crucial to obtain a $(\pi,\pi)$-AM. It also reduces the critical interaction for the formation of moments on the O atoms, i.e. the $(0,0)$-AM. 
Parameters used, chosen to be realistic for cuprates~\cite{Sheshadri2023, Weber2012, Weber_2012_arxiv}, are:
$t_{pd} = 1.3$ eV, $t_{pp} = 0.6$ eV, and $\epsilon_p = 2.2$ eV; 
order parameter threshold $10^{-2}$, tol $10^{-5}$ using a solver (see main text). Note that reasonable values of $U_d$ in cuprates are from 7 eV to 10.5 eV while values from $U_p$ are 4 eV to 6 eV~\cite{Sheshadri2023}.}
\label{fig:MFT_phases1}
\end{figure}

In general, the presence of AM in the cuprate planes requires a magnetic moment on the oxygen, thus, not full but only partial occupation, and a finite Hubbard repulsion $U_p$.
It naturally appears in the negative charge transfer limit $\epsilon_p <0$ at half-filling, when the oxygen orbitals are above the Cu d-orbitals and their partial occupation then enables oxygen AFM once a small Hubbard repulsion $U_p$ is present. Consequently $(0,0)$-AM sets in. 

Interestingly, AM can also become stable in the far more realistic {\it positive} charge transfer regime $\epsilon_p >0$.
It has been shown that already for a low but still realistic charge transfer energy of $\epsilon_p = 2.2$ eV~\footnote{And for infinitely large copper repulsion $U_d$ and realistic oxygen-oxygen hopping $t_{pp} = 0.6$ eV.} the magnetic moment calculated for the Emery model at half-filling has a 37\% contribution from oxygen electrons~\cite{Jefferson1992}. This nonzero occupancy can be further enhanced by lowering $\epsilon_p$ {\it or} by increasing hole doping $\delta$ (see below).
Naturally, a finite occupancy of oxygen orbitals only triggers AM once the Hubbard repulsion $U_p$ is strong enough. Therefore, in the following we explore $U_p$ dependence of the phase diagram of the Emery model in the realistic positive charge transfer limit. 
%
%
To this end,
we elucidate $U_p$-driven oxygen AFM within a standard MFT. 

In the MFT we first split
the Hamiltonian $\mathcal{H}$, Eq.~\eqref{eq:hamiltonian}, into a non-interacting part $\mathcal{H}_0$ and an interacting part $\mathcal{H}_U$, such that $\mathcal{H} = \mathcal{H}_0 + \mathcal{H}_U$.
The non-interacting Hamiltonian $\mathcal{H}_0$ can be diagonalized by Fourier transformation, $c_{k \mu \sigma} = N^{1/2} \sum_i e^{-\mathrm{i} k R_{i \mu}} c_{i \mu \sigma}$ and $R_{i \mu}$ is the position of orbital $\mu$ in the unit cell $i$. We obtain $\mathcal{H}_0 = \sum_{k,\sigma} (c_{k d \sigma},c_{k x \sigma},c_{k y \sigma})^\dag h_0(k) (c_{k d \sigma},c_{k x \sigma},c_{k y \sigma})$ where
\begin{align}
    h_0(k) = 
    \left(
    \begin{matrix}
        \epsilon_d& 2 \mathrm{i} t_{pd} s_x & -2 \mathrm{i} t_{pd} s_y\\
        -2 \mathrm{i} t_{pd} s_x&\epsilon_p&
        4 t_{pp} s_x s_y\\
        2 \mathrm{i} t_{pd} s_y&4 t_{pp} s_x s_y&\epsilon_p
    \end{matrix}
    \right)
\end{align}
with $s_{x(y)} = \sin\left(k_{x(y)}/2 \right)$.
We focus on $(0,0)$ and $(\pi,\pi)$ instabilities and introduce two sublattices $\lambda = A,B$ labeling the two copper atoms in the two copper unit cell. We then combine the fermion operators in a  twelve-component vector $\vec{\Psi}_{\vec{k}}$ (with indices explicitly written as $\Psi_{\vec{k}\mu\lambda \sigma}$).

We define the mean fields $m_d = \sum_i (-1)^{i} \langle S^z_{i d} \rangle/N$,
i.e. the staggered magnetization of the Cu atoms, and $m_p = \sum_i \langle (S^z_{i x} - S^z_{i y}) \rangle/N$, the intra-unit cell staggered magnetization of the $p$ orbitals. Note that by default there are two $p$ orbitals with opposite magnetization in the Wigner--Seitz cell. To complete our ansatz, we also need to account for an interaction-induced change of the relative occupation of the orbitals. By proper renormalization of the chemical potential, it can be incorporated by introducing the third mean field $n_d = \sum_{i,\sigma} \langle c^\dag_{i d \sigma} c_{i d \sigma} \rangle/N$ as the filling of the $d$ orbitals in analogy to the overall filling $n$. Decoupling the Hamiltonian in the charge channel, we find 
\begin{align}
    \mathcal{H}_U =& 
    \frac{U_d N}{8} \left(m_d^2-n_d^2\right) + \frac{U_p N}{16} \left(m_d^2-(n-n_d)^2\right)
    \nonumber \\
    &+
    \sum_{k,\lambda,\sigma} \frac{U_d}{4} \left(n_d -(-1)^{\lambda +\sigma} m_d  \right) c^\dag_{k d \lambda \sigma} c_{k d \lambda \sigma} 
    \nonumber \\
    &-  
    \frac{U_p}{8} (-1)^{\sigma} m_p
    \left(c^\dag_{k x \lambda \sigma} c_{k x \lambda \sigma}
    - c^\dag_{k y \lambda \sigma} c_{k y \lambda \sigma} \right).
\end{align}

\begin{figure}
\includegraphics[width=\linewidth]{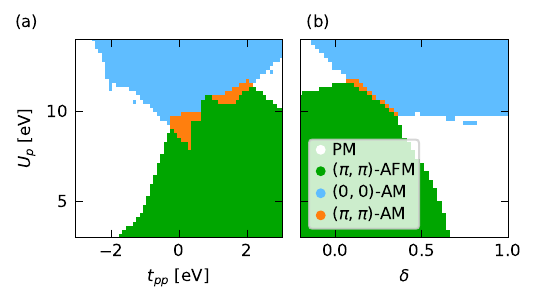}
\caption{MFT phase diagram as function of  oxygen--oxygen hopping $t_{pp}$ at doping $\delta=0.5$, panel (a), and as function of hole doping $\delta$ at $t_{pp}=0.6$ eV, panel (b).
[
chosen parameters, realistic for cuprates~\cite{Sheshadri2023, Weber2012, Weber_2012_arxiv}, are:
$t_{pd} = 1.3$ eV, $\epsilon_p = 2.2$ eV, $U_d=8$ eV;
order parameter threshold $10^{-2}$, tol $10^{-5}$ using a solver (see main text)].}
\label{fig:MFT_phases2}
\end{figure}

The resulting MFT Hamiltonian $\sum_{\vec{k}} \Psi_{\vec{k}}^\dag h(\vec{k}) \Psi_{\vec{k}} + E_0$
see Appendix~\ref{app:MF} Eq.~\eqref{eq:MFT_hamiltonian_matrix}, 
is a non-interacting Hamiltonian. The $12\times 12$ Bloch Hamiltonian $ h(\vec{k})$ can be efficiently diagonalized for each momentum to obtain the eigenenergies $\epsilon_m(\vec{k})$ and the Bloch eigenstates $|u_m(\vec{k})\rangle$ with band index $m$. Note that $H$ is block-diagonal in spin, but in contrast to the conventional AFM spin density mean-field solution~\cite{knolle2011multiorbital} the two blocks are explicitly spin-dependent. We have solved the mean-field equations self-consistently for fixed filling $n$, i.e. for $\vec{\mathcal{M}} = (m_d,m_p,n_d,\mu)$ we calculated $\vec{\mathcal{M}}_r$ from $H(\vec{\mathcal{M}}_{r-1})$ until $r=100$, and then applied a python in-build solver which solves the fixpoint equation self-consistently.

%

The resultant MFT results are displayed in Fig.~\ref{fig:MFT_phases1} and Fig.~\ref{fig:MFT_phases2} where we find wide regions with both types of AM states. The effect of different parameters is summarized in Table~\ref{tab:results}. First, we observe that the key driving term is the interaction $U_p$ on the oxygen orbital, which in conjunction with a direct hopping $t_{pp}$ generates a strong AFM exchange between the oxygen moments. However, in order to avoid a full occupation of the $p$-orbitals, non-zero doping is required or a raise of the on-site energies $\epsilon_p$. At zero doping the single hole is always located entirely on the $d$-orbital or entirely on the $p$-orbital, no $(\pi,\pi)$-AM forms. Finite hole doping allows to distribute the additional hole charge between the otherwise competing orbitals which becomes visible by the formation of a $(\pi,\pi)$-AM state, at an increasing critical $U_d$, and a decreasing critical $U_p$. Hole doping therefore strengthens the AM phases and weakens the AFM phase, see Fig.~\ref{fig:MFT_phases2}~(b).

The fact that oxygen and copper are competing for hole charge in order to form a magnetic moment is also evidenced by the observation that the critical $U_d$ for AFM increases when $U_p$ is increased and  vice versa, the critical $U_p$ for $(0,0)$-AM increases when $U_d$ is increased.

Beyond a basic direct exchange picture, the dependence on $t_{pp}$ indicates the importance of a copper-mediated oxygen--oxygen superexchange. The MFT results, see Fig.~\ref{fig:MFT_phases2}~(a), show that oxygen moments form even for $t_{pp}=0$ and the lowest critical $U_p$ is found for small positive $t_{pp}$. Note that the phase diagram is not symmetric under sign changes of $t_{pp}$, because for negative $t_{pp}$ the FS has an increased contribution of the $p$-orbitals and has not only $d$-orbital character. 

\section{AM driven by oxygen-oxygen hopping $t_{pp}$}
%
%
While the MFT has shown that AM states can be stabilized, we next explore the phase diagram within ED calculations. We confirmed the $U_p$-driven AM observed in MFT, but interestingly find another route to intra-unit cell oxygen AFM that does not depend on $U_p$. Instead, it can be triggered by  oxygen-oxygen hopping $t_{pp}$. As this subtle effect depends on the nature of the correlated many-body wave function, it appears in ED but not simple MFT and can be accounted for within cell perturbation theory.

%

%
%
Using the three-orbital Emery model Hamiltonian [Eq.~\eqref{eq:hamiltonian}] we performed ED for a cluster of four Cu and eight O with periodic boundary conditions at zero doping (4 holes) as well as $\delta$ = 0.25 (5 holes), shown in Fig.~\ref{fig:MFT_phases1} (b). We used $U_d$ = 8 eV and $U_p$ = 4 eV. We note that for our small ED cluster the resulting ground states do not exhibit a finite magnetization since they cannot spontaneously break symmetries. Instead, we characterize the phases by measuring the spin-spin correlations $\langle \mathbf{S}_i \cdot \mathbf{S}_j \rangle$ for the nearest neighbour Cu-Cu and O-O sites (see Appendix~\ref{app:ED}) to diagnose the ($\pi$, $\pi$)-AFM, (0,0)-AM, and ($\pi$, $\pi$)-AM states shown in Fig.~\ref{fig1}. If spin correlations are incompatible with our ansatzes, phases are labeled as {\it others}. 

\begin{figure}
\includegraphics[angle=0,width=\linewidth]{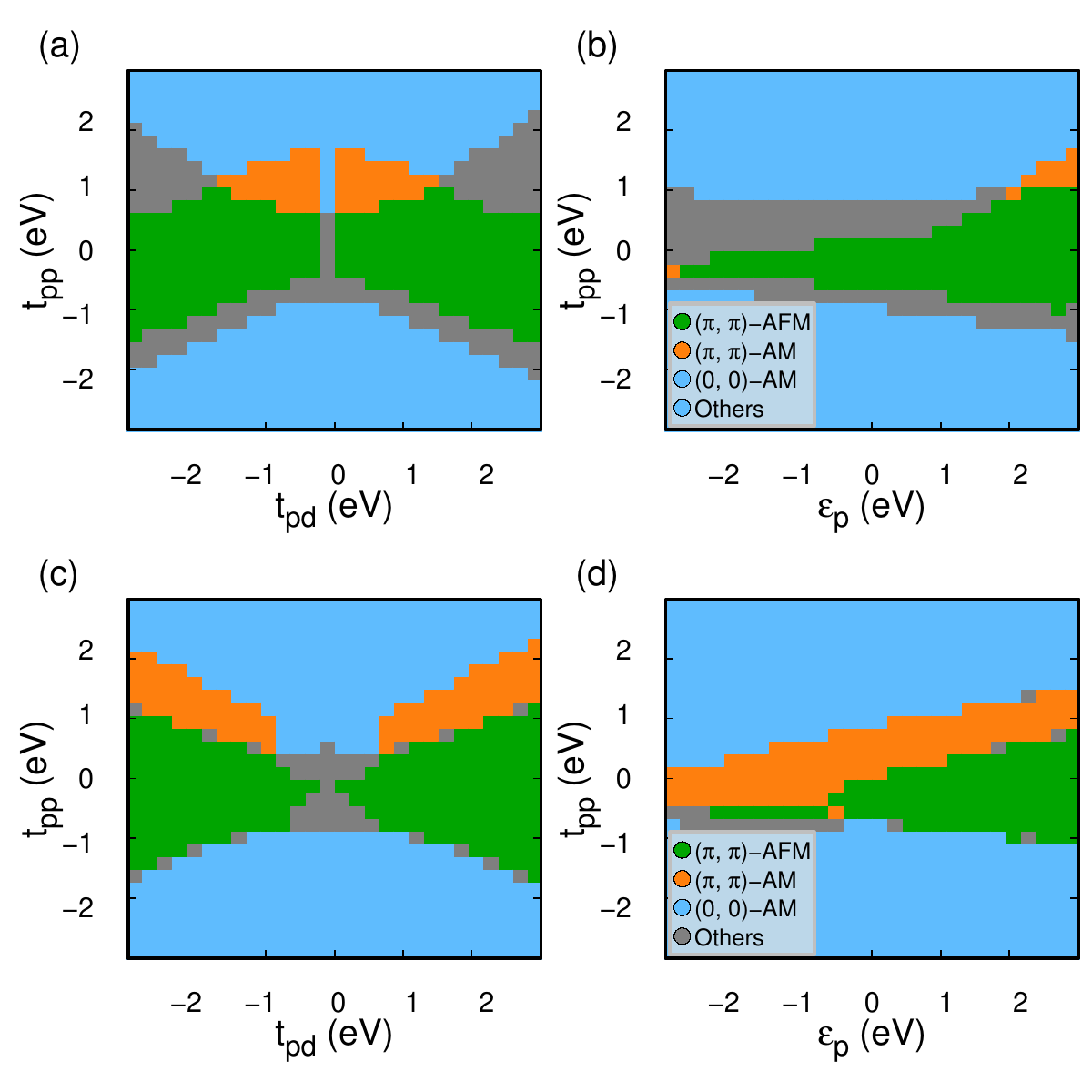}
\caption{Phase diagram from ED  ($U_d$ = 8 eV, $U_p$ = 4 eV) for zero doping (4 holes) as a function of (a) $t_{pd}$ and $t_{pp}$ ($\epsilon_p$ = 2.2 eV, ) as well as (b) $\epsilon_p$ and $t_{pp}$ ($t_{pd}$ = 1.3 eV). (c) and (d) are the corresponding cases for doping $\delta$ = 0.25 (5 holes).
All fixed parameters correspond to realistic values for cuprates~\cite{Sheshadri2023, Weber2012, Weber_2012_arxiv}.
}
\label{fig:structed}
\end{figure}

The obtained phase diagram for zero doping is displayed in Fig.~\ref{fig:structed} (a) for $t_{pd}$ and $t_{pp}$ ($\epsilon_p$ = 2.2 eV) as well as (b) $\epsilon_p$ and $t_{pp}$ ($t_{pd}$ = 1.3 eV). (c) and (d) are corresponding results for $\delta$ = 0.25. Similar to the MFT results, doping could enlarge the $(\pi, \pi)$-AM region. The results differ in two key aspects from the MFT results. First, we can identify $t_{pp}$ as the main driving term of AM, see Fig.~\ref{fig:structed}~(a). Second, and in strong contrast to the MFT result, the formation of oxygen moments depends only weakly on $U_p$ . It points towards a different mechanism driven by the hopping $t_{pp}$ instead of the interaction $U_p$, which we discuss below.


The sign of $t_{pd}$ does not affect the phase diagram while $\epsilon_p$ and $t_{pp}$ are essential. When the magnitude of $t_{pp}$ is small, the phase is a ($\pi$, $\pi$)-AFM, while increasing $t_{pp}$ can enhance AFM correlations between oxygens and reduce them between coppers, leading to the transition to ($\pi$, $\pi$)-AM, and then to (0,0)-AM. When $t_{pp}$ is fixed, smaller (as negative as possible) $\epsilon_p$ increases the hole contribution of oxygen, inducing AM correlations.

The main result obtained by ED is that in the limit of realistic  charge transfer energy $\epsilon_p$, oxygen-copper hopping $t_{pd}$, and oxygen repulsion $U_p$, AFM correlations between oxygens already appear at half-filling, i.e. one hole per CuO$_2$ unit cell, for only slightly enhanced values of $t_{pp}$. Moreover, in this regime oxygen magnetism is independent of the value of the oxygen repulsion $U_p$ -- instead it is triggered by the oxygen-oxygen hopping $t_{pp}$.  

In Table~\ref{tab:results} we summarize the parameter regimes supporting AFM correlations on the oxygens (and therefore AM correlations) obtained from our ED calculations compared to MFT.  


\begin{table}
    \centering
    \caption{Summary of the parameter regimes supporting AFM correlations on the {\it oxygens}  as obtained from ED compared to MFT. 
    }    
    \begin{ruledtabular}
    \begin{tabular}{c|c|c}
         &  MFT & ED\\
         \hline
        $t_{pp}$ & $\approx 0$ or small but positive & $|t_{pp}|$ large\\
        $U_d$ & small & weak dependence\\
        $U_p$ &large $U_p$ crucial& weak dependence\\
        $\delta$ & $\delta>0$ crucial & $\delta>0$ \\
        $\epsilon_p$ &  \multicolumn{2}{c}{small (as negative as possible)}
    \end{tabular}
    \end{ruledtabular}
    \label{tab:results}
\end{table}

The ED findings that AFM correlations on oxygens can strongly depend on $t_{pp}$ but only weakly on $U_p$ are rather counterintuitive. 
Interestingly, however, for the $(0,0)$--AM case, this can be partially rationalized by noting that there exist spin-exchange processes between magnetic moments on oxygens that do {\it not} depend on $U_p$. In fact, in the limit of large $|\epsilon_p|, U_p, U_d$ and no occupancy on copper sites, following~\cite{Kaushal2024}, we deduce that the nearest-neighbor spin exchange between oxygens has three lowest-order contributions: (i) a dominant second-order term proportional to $t_{pp}^2 / U_p$, (ii) third-order terms proportional to $t_{pd}^2 t_{pp} / \epsilon_p^2$ and $t_{pd}^2 t_{pp} / (|\epsilon_p| U_p)$, and (iii) fourth-order terms proportional to $t_{pd}^4 / (\epsilon_p^2 (U - 2\epsilon_p))$ and $t_{pd}^4 / (\epsilon_p^2 (\epsilon_p + U_p))$. Hence, in the strong-coupling perturbative limit, there are indeed spin-exchange processes whose strength does not depend on $U_p$. Naturally, what remains to be better understood is how such a strong-coupling result persists when $\epsilon_p$ is {\it not} large and negative, and the assumption of empty copper sites no longer holds. We leave this investigation for future work.

In contrast, the $(\pi, \pi)$--AM case presents a far more complex situation, as it requires understanding how the two seemingly competing AFM orders—on oxygen and copper—can coexist.
In the following we provide a qualitative picture motivated by a cell perturbation theory~\cite{Jefferson1992}. Following Ref. ~\cite{Jefferson1992} we can rewrite the charge transfer model Eq.\eqref{eq:hamiltonian} in a different oxygen basis.  We define the bonding and antibonding combinations of the four nearest neighbor oxygen
orbitals (labeled by $L,R,T,B$) surrounding the copper $d$ orbital and obtain
$|\alpha \rangle 
\equiv (|p_{x R} \rangle -|p_{y T} \rangle 
 - |p_{x L} \rangle+ |p_{y B} \rangle)/2$
 and $|\beta \rangle 
\equiv (|p_{x R} \rangle +|p_{y T} \rangle 
 - |p_{x L} \rangle- |p_{y B} \rangle)/2$. Interestingly, we can discard the oxygen repulsion $U_p$ completely such that the Hamiltonian takes the form
\begin{align} \label{eq:JHamiltonian}
    \mathcal{H}=& 
%
    \epsilon_p
    \sum_{i, \sigma} (\alpha^\dag_{i  \sigma} \alpha_{i, \sigma} + \beta^\dag_{i  \sigma} \beta_{i \sigma} ) 
    + 2 \sum_{i, j (i),  \sigma}  \Big[ t_{pd} \mu (j) 
    c^\dag_{i d \sigma} \alpha_{j \sigma}
    \nonumber \\
    &+  t_{pp} \nu(j)
    ( \alpha^\dag_{i \sigma} \alpha_{j \sigma}
    -\beta^\dag_{i \sigma} \beta_{j \sigma}
    ) + t_{pp} \chi(j)
    \alpha^\dag_{i \sigma} \beta_{j \sigma} + {\rm H.c.} \Big]  \nonumber \\
    &+  \sum_{i} U n_{id\uparrow}n_{id\downarrow},  
\end{align}
where
$\alpha^\dag_{i, \sigma}$ 
and $\beta^\dag_{i, \sigma}$ create holes in orthogonalised orbitals
$\alpha$ and $\beta$ centered around copper site $i$
and $\mu(j), \nu(j), \chi(j)$ are the orthogonalisation factors that strongly depend on the distance between the CuO$_4$ cells $j$,
see ~\cite{Jefferson1992} and Table~\ref{tab:orthofactors} in Appendix~\ref{sec:orthofactors}.
The crucial point is that the hybridization strength $\chi(j)$ does not have an on-site component 
[i.e. $\chi(0) = 0$],
though it couples nearest-neighbor orbitals $\alpha$ and $\beta$.

In most treatments of the charge transfer model Eq.\eqref{eq:JHamiltonian} the $\beta$ orbitals are neglected leading to a two-band model. This, {\it inter alia}, gives rise to the well-studied AFM superexchange processes and the Zhang-Rice singlets upon doping the cuprates \cite{Zhang1988}. However, our key observation is that neglecting the $\beta$ orbitals is no longer justified once $|t_{pp}|$ is  larger than assumed in most cuprate works and once the $\beta$ orbital is not fully occupied. It is the nearest neighbor hybridization between oxygen orbitals that induces oxygen AFM.

Then a simple picture emerges best explained in a minimal toy-model. From 
Eq.\eqref{eq:JHamiltonian} we concentrate on two plaquettes (one of each A and B sublattices) and two holes. First, for infinite Coulomb repulsion on the copper $d$ orbital strong super exchange leads to copper AFM~\cite{Jefferson1992}, e.g. plaquette A spin up
in the $d$ orbital and plaquette B spin down. 
%
Next, the standard local hybridization 
between $d$ and $\alpha$ orbitals leads to  AFM correlations between plaquettes admixing oxygen states to the magnetic moment~\cite{Jefferson1992}. Going beyond this picture, we  take into account the nearest neigbor $\alpha,\beta$ 
hybridisation.
An up spin in $\alpha$ orbitals of site A induces an up spin in $\beta$ orbitals at site B and vice versa. As a result, the spin of bonding/antibonding oxygen orbitals is anti-aligned in each plaquette (thus aligned between neighboring plaquettes). 
The hybridisation 
between the nearest neigbor $\alpha,\beta$ orbitals
is directly proportional to $t_{pp}$ and when calculating spin correlations in this state between the original oxygen ligands in the unit cell they are indeed AFM. 


\section{AM driven by large doping in a material candidate}
%
%
\begin{figure}
\includegraphics[angle=0,width=\linewidth]{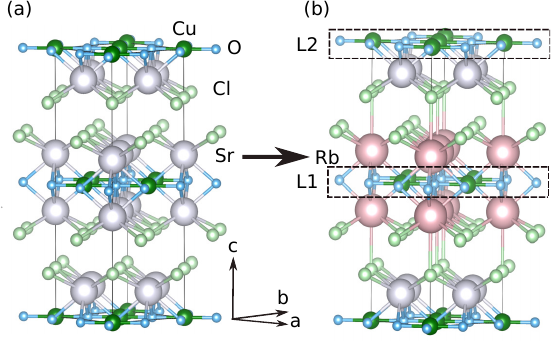}
\caption{Crystal structure of (a) Sr$_2$CuO$_2$Cl$_2$ and (b) SrRbCuO$_2$Cl$_2$. L1 and L2 label two CuO$_2$ layers.}
\label{fig:structsrrb}
\end{figure}
Finding large values of $U_p$ and $t_{pp}$ in cuprate materials is not straightforward, therefore, we considered a few candidates for AM using {\it ab initio} DFT-based calculations. The first candidate is the widely studied cuprate \LCO, where it is difficult to obtain in-plane oxygen moments since the extra holes from doping prefer to stay in apices O locations. We therefore turn our attention to the compound \SCOCl~\cite{Miller1990}, where the out-of-plane oxygen ions at the apices of the CuO$_6$ octahedra are replaced by Cl. In addition, a smaller $\epsilon_p$ has also been  proposed, {\it cf}.~\cite{Weber2012} 
and Table I available in its arXiv version~\cite{Weber_2012_arxiv}.

In Fig.~\ref{fig:structsrrb} (a) we show the tetragonal structure of  \SCOCl\ with space group $I4/mmm$. Unlike \LCO, no long-range tilting/rotation of the CuO$_4$Cl$_2$ octahedra is present. The structure leads to a very small magnetic anisotropy~\cite{Cuccoli2003, Suh1996, Katsumata2001, Wells1995} and very weak interlayer coupling~\cite{Wells1995}, making \SCOCl~a better two-dimensional AFM insulator~\cite{Wells1995,Yaresko2002, Torre2021}.  Both, inelastic neutron and light scattering measurements found a high-energy continuum of magnetic fluctuations~\cite{Plumb2014} and multiple-magnon excitations~\cite{Betto2021}. Similar to \LCO, the valence of the Cu atoms is +2, with one hole per copper atom. 
In order to have oxygen magnetic moments, we replaced half of the Sr atoms by Rb so that the system \SRCOCl~[See Fig.~\ref{fig:structsrrb} (b)] is hole doped, and becomes metallic. The CuO$_4$ layer neighbor to Rb (Sr) is labeled as L1 (L2). These two  layers capture different dopings, as discussed in the next paragraph. The structure was fully relaxed with GGA using the Vienna {\it ab initio} simulation package (VASP)~\cite{Kresse1996,Hafner2008}. Total energies for various spin configurations and corresponding band structures were calculated with GGA+$U$.  We considered contributions of the Coulomb repulsion~\cite{Dudarev1998}, $U_{\rm eff}$ = $U_d- J_H$ = 8 eV ($U_d = 8$ eV, $J_H = 0$ eV) for Cu and $U_p$ = $J_H$ = 0~eV for O following~\cite{Sterling2021} within GGA+$U$.  Note that in the Dudarev's implementation of GGA+$U$, U and $J_H$ are implemented as $U_{\rm eff}$ = $U_d- J_H$, therefore the combinations of $U_d$ and $J_H$ with the same $U_{\rm eff}$ would not change the results. Here we observe that In GGA+$U$, a finite $U$ stabilizes the magnetic moments of Cu.

\begin{figure}
\includegraphics[angle=0,width=\linewidth]{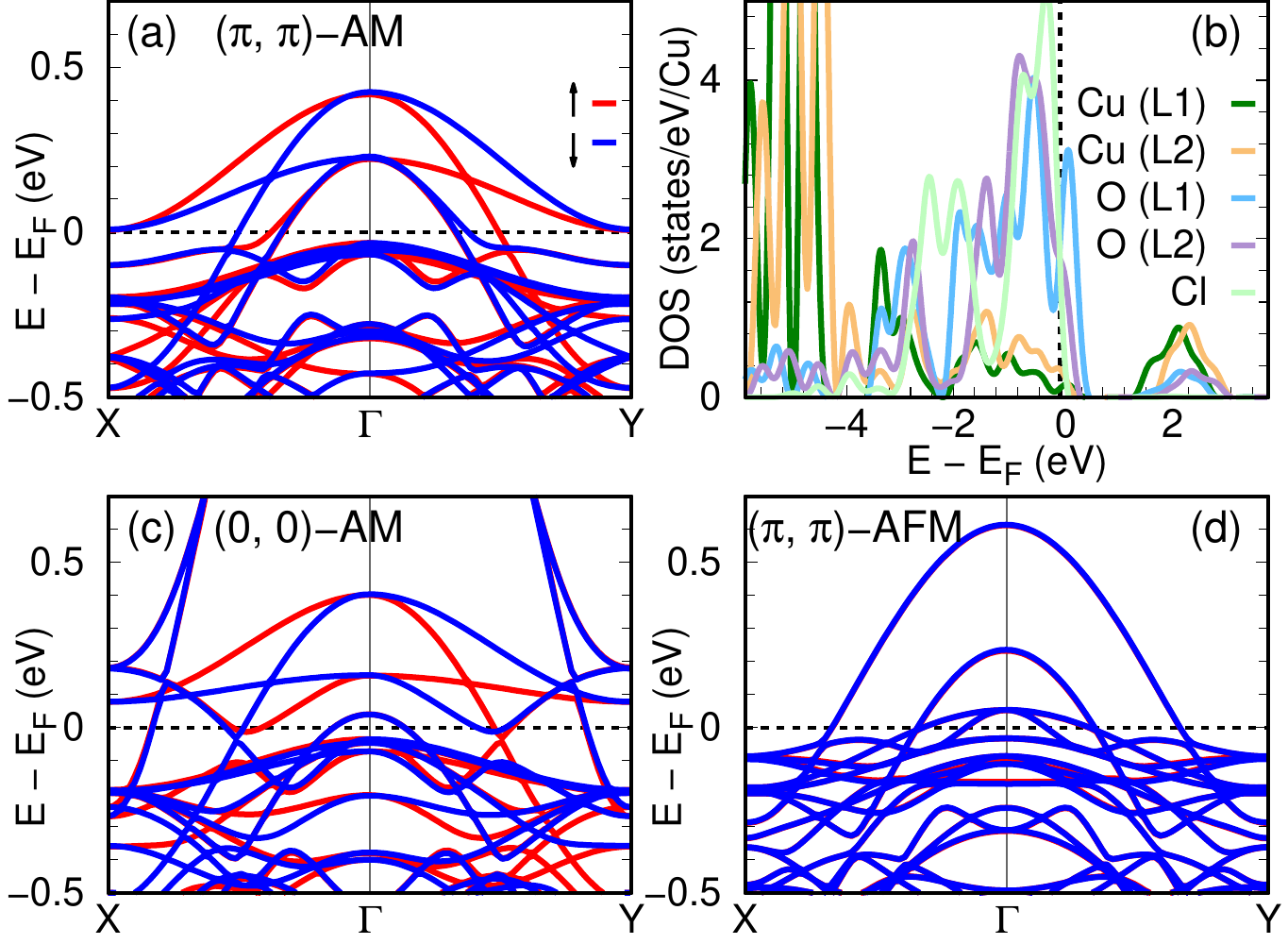}
\caption{(a) Band structure and (b) density of states for \SRCOCl~in the ($\pi$, $\pi$)-AM magnetic order within GGA+$U$ from DFT as shown in Fig.~\ref{fig1}. Band structure in the (c) (0, 0)-AM magnetic order and (d) ($\pi$, $\pi$)-AFM order are also shown. Colors red (spin $\uparrow$) and blue (spin $\downarrow$) indicate the up bands and down states, respectively. The AM ordering is clear from the spin splitting along $\Gamma$-X and $\Gamma$-Y.}
\label{fig:DFTbanddos}
\end{figure}

By comparing the energy for various magnetic configurations;  Cu AFM order,  (${\pi, \pi}$)-AM order, and  (0, 0)-AM order, we find that (${\pi, \pi}$)-AM order is 241.5 meV lower than (0, 0)-AM and 53.9 meV lower than ($\pi$, $\pi$)-AFM. The magnetic moments are 0.6 $\mu_{\rm B}$ for Cu, similar to \LCO~\cite{Lane2018}. However, in contrast to \LCO, which has a small oxygen moment on the apical oxygen~\cite{Lane2018} but no doping-induced oxygen magnetism in the CuO$_2$ plane, in \SRCOCl\ the magnetic moments are 0.4 $\mu_{\rm B}$ for oxygen. We integrated the partial density of states for Cu and O and find that the additional four holes from doping are mostly localized at the four O in the L1 layer shown in Fig.~\ref{fig:DFTbanddos} (b) with  doping  $\delta$ = 2 (3 electrons, 3 holes per CuO$_2$). 

The resulting band structure within GGA+$U$ ($U_d$ = 8 eV, $U_p$ = 0 eV) obtained from DFT is displayed in Fig.~\ref{fig:DFTbanddos}  for
($\pi$, $\pi$)-AM order, (0,0)-AM order, and ($\pi$, $\pi$)-AFM order. The bands have opposite spin-splitting sign along $\Gamma$-X and $\Gamma$-Y. 
We stress that, different from the case of \LCO, where AM originates from CuO$_6$ octahedra distortions, in  \SRCOCl\ such distortions are absent and AM  arises from oxygen moments.

\section{Conclusion}
\begin{figure}
    \centering
    \includegraphics[width=\columnwidth]{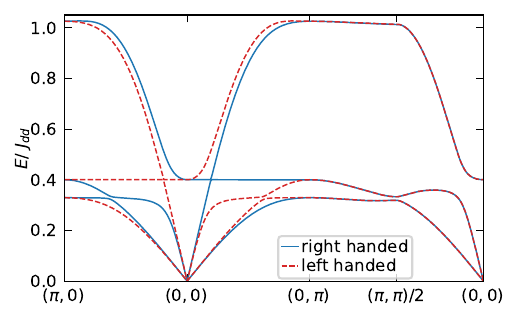}
    \caption{Magnon spectrum along a high symmetry path through the magnetic Brillouin zone, determined from linear spin wave theory of an effective Heisenberg-like spin only model with moments on the oxygens and copper atoms for the $(\pi,\pi)$-AM state, see Eq.~\eqref{eq:H_heisenberg} in Appendix~\ref{app:mag}. 
    The different chiralities (left and right precessing) are distinguished by color and linestyle. Note, we have used an artificially large splitting for presentation. 
    [$J_{pd}=0.5 J_{dd}$ and $J_{pp}=0.4 J_{dd}$]}
    \label{fig:magnon_spectrum}
\end{figure}
We have studied $d$-wave magnetism in cuprates induced by magnetic moments on the oxygen atoms. Focusing on the three-orbital Emery model we employed Hartree-Fock MFT, ED of finite clusters, and cell perturbation theory to uncover distinct mechanisms stabilizing AFM correlations between oxygen moments. Two different AM states can be stabilized. First, a $(0,0)$-AM with AFM ordered moments only on the oxygen sites, such that the magnetic unit cell is identical with the crystal unit cell. The Fermi surface is characterized by a large spin splitting of the formerly $(\pi,\pi)$ centered pocket into open Fermi surfaces along the crystallographic axis, see Fig.~\ref{fig1}. Interestingly, this intra-unit cell magnetic order has been  proposed previously to explain inelastic neutron scattering data~\cite{fauque2006magnetic,li2008unusual}. Secondly, we find a $(\pi,\pi)$-AM with AFM order of the oxygens as above coexisting with the standard $(\pi,\pi)$ magnetic order on the copper atoms. Here, we observe a Fermi surface with two closed pockets which are additionally weakly spin split by opposite tilting, see Fig.~\ref{fig1}.

Our work shows that while AM states in cuprates can be stabilized, the required parameter regimes for $t_{pp}$ and $U_p$ are challenging to obtain for typical cuprates. The exact form of the phase boundaries is method dependent, e.g. MFT favoring large $U_p$ while our ED of small clusters points to the relevance of $t_{pp}$. Despite these challenges, we identify parameters which are potentially relevant for certain cuprate families. Based on our observations, we then identify three distinct mechanisms potentially realizing AM in cuprates: (i) a direct exchange between the oxygens requiring a large oxygen interaction $U_p$, (ii) strong charge transfer, i.e., $\epsilon_p$ small, bringing the $p$-oxygen orbital closer to the Fermi energy, and (iii) a large oxygen--oxygen hopping $t_{pp}$ effectively such that the magnetic order of the copper atoms induces magnetic order on oxygen sites.
We note that {\it ab-initio} calculations of Ref.~\cite{Weber2012},
and in particular Table I available in its arXiv version~\cite{Weber_2012_arxiv}, suggest that a number of cuprates have surprisingly low charge transfer energy $\epsilon_p < 2$eV
and a large oxygen-oxygen hopping $t_{pp} \sim 0.6-0.7$ eV. 
Further indications of the substantial values of the oxygen-oxygen hopping in the cuprates are discussed in e.g. Refs.~\cite{Pavarini2001, Barii2022, Martinelli2024}.

Beyond  our proof of principle calculations on the minimal Emery model, we also investigated the realization of AM in a novel candidate material \SRCOCl\ based on the well-known existing system \SCOCl. The DFT results suggest the (${\pi, \pi}$)-AM order to be the most stable one emerging from AFM correlations between oxygen moments.
Since the AM in our case is induced by magnetism on the $p$ orbitals, detailed DFT calculations for
each material are required. As this is computationally very involved, we leave it for a next
study to do a database materials. 

The prospect of AM in doped cuprates raises the question on how it would manifest in experiments. Apart from the standard spin transport signatures predicted for AM~\cite{Libor2022a,Libor2022,gonzalez2024},
 angle resolved photoemission spectroscopy (ARPES) or spin ARPES~\cite{Osumi2024} could directly probe the spin splitting of the Fermi surface. Alternatively, the split FS can be detected by quantum oscillation measurements which have a long history in underdoped cuprates \cite{doiron2007quantum,sebastian2012towards}. Interestingly, since Zeeman coupling will lead to different areas for the spin-polarized FSs, the temperature dependence of QO can be unconventional~\cite{Li2024,mccollam2005anomalous} and new low frequency oscillations could appear~\cite{leeb2023theory,leeb2024field}. 
Most promising would be to revisit previous inelastic neutron scattering experiments~\cite{fauque2006magnetic,li2008unusual}. Magnon spectra of AM display chirality split magnons away from high symmetry points. The splitting is potentially small and requires high resolution. In Fig.~\ref{fig:magnon_spectrum} we show predictions for magnon spectra with chirality split magnon bands. Note, we treated a simple Heisenberg like model of spins on copper and oxygen sites, see Appendix~\ref{app:mag} for details, and the optical magnon modes from intra-unit cell excitations are strongly split, but because of the weak oxygen moment might be hard to observe. 
Finally, AM in cuprates or other related compounds would have important ramifications for superconductivity. AM have been shown to lead to unconventional phenomena~\cite{Ouassou2023, Sun2023,Ferrari2024,Shao2021, Rafael2021} such as pair density states (PDW) with finite momentum pairing~\cite{zhang2024finite,zhu2023topological,chakraborty2024zero,sim2024pair}.
Especially the $(0,0)$-AM offers -- due to its large spin-splitting and absence of magnetism for the copper sites --  an interesting platform for realizing PDWs. Searching for signatures thereof in experiment and understanding the microscopic mechanism for its realization in realistic settings will be important directions for future research. 

{\it Note added.} Upon completion of this work, we became aware of two related independent works. D\"urrnagel {
\it et al.} \cite{Ref-Thomale} explored an itinerant AM phase transition emanating from an electronic model on a related Lieb lattice~\cite{Brekke2023}, where AM originates from sublattice interference of the enlarged unit cell. Kaushal {\it et al.} \cite{Kaushal2024} studied itinerant AM in a Lieb lattice Hubbard model for a different parameter regime of an `anti-CuO$_2$' lattice. 

\section*{Acknowledgments}
We thank Mark Fisher, Kemp Plumb, Alexander Mook and Igor I. Mazin for insightful discussions. Y. L. acknowledges support from the Alexander von Humboldt Foundation through a postdoctoral Humboldt fellowship and Fundamental Research Funds for the Central Universities (Grant No. xzy012023051). V. L. acknowledges support from the Studienstiftung des deutschen Volkes. J. K. acknowledges support from the Imperial-TUM flagship partnership. R.V. acknowledges support by the Deutsche Forschungsgemeinschaft (DFG, German Research Foundation) for funding through Project No. TRR 288 --- 422213477 (project A05, B05) and Project No. VA 117/23-1 --- 509751747.
K.W. acknowledges the support of National Science Centre in Poland under project no.
2024/55/B/ST3/03144.
Collaboration between J.K., L. V. and K.W. was supported by the `Tandems for Excellence: Visiting Researchers Programme' of the University of Warsaw.

\appendix

\section{Mean-field theory}
\label{app:MF}
The $12\times 12$ Hamiltonian is block-diagonal in spin. We consider the ordered basis
\begin{align}
    \psi_{\vec{k}} = (& c_{\vec{k}dA \uparrow},c_{\vec{k}xA \uparrow},c_{\vec{k}yA \uparrow},c_{\vec{k}dB \uparrow},c_{\vec{k}xB \uparrow},c_{\vec{k}yB \uparrow},
    \nonumber \\ & 
    c_{\vec{k}dA \downarrow},c_{\vec{k}xA \downarrow},c_{\vec{k}yA \downarrow},c_{\vec{k}dB \downarrow},c_{\vec{k}xB \downarrow},c_{\vec{k}yB \downarrow}
    )^\trans.
\end{align}
The matrix elements are 
\begin{align}
    h(\vec{k}) =& \1 \otimes \begin{pmatrix} h_{AA}(\vec{k}) & h_{AB}(\vec{k}) \\ h_{AB}(\vec{k})^\dag & h_{AA}(\vec{k})\end{pmatrix}
    \nonumber\\&
    - \frac{U_d m_d}{4} \tau^z \otimes \tau^z \otimes \begin{pmatrix}1&0&0\\0&0&0\\0&0&0\end{pmatrix}
    \nonumber\\&
    + \frac{U_d n_d}{4} \1 \otimes \1 \otimes \begin{pmatrix}1&0&0\\0&0&0\\0&0&0\end{pmatrix}
    \nonumber\\&
    - \frac{U_p m_p}{8} \tau^z \otimes \1 \otimes \begin{pmatrix}0&0&0\\0&1&0\\0&0&-1\end{pmatrix}
    \nonumber\\&
    + \frac{U_p (n-n_d)}{8} \1 \otimes \1 \otimes \begin{pmatrix}0&0&0\\0&1&0\\0&0&1\end{pmatrix}
    \\
    h_{AA}(\vec{k}) =&
    \begin{pmatrix}
        0 & -t_{pd} \e^{-\ii k_x/2} & t_{pd} \e^{-\ii k_y/2} \\
        -t_{pd} \e^{\ii k_x/2} & 0 & 2 t_{pp} \cos\left(\frac{k_x-k_y}{2}\right) \\
        t_{pd} \e^{\ii k_y/2} & 2 t_{pp} \cos\left(\frac{k_x-k_y}{2}\right) & 0 
    \end{pmatrix}
    \\
    h_{AB}(\vec{k}) =&
    \begin{pmatrix}
        0 & t_{pd} \e^{-\ii k_x/2} & -t_{pd} \e^{-\ii k_y/2} \\
        t_{pd} \e^{\ii k_x/2} & 0 & -2 t_{pp} \cos\left(\frac{k_x+k_y}{2}\right) \\
        -t_{pd} \e^{\ii k_y/2} & -2 t_{pp} \cos\left(\frac{k_x+k_y}{2}\right) & 0 
    \end{pmatrix}
    \label{eq:MFT_hamiltonian_matrix}
\end{align}
and $\tau^z$ is the Pauli-$z$ matrix.

\section{Spin-Spin correlations from ED}
\label{app:ED}
\begin{figure}
\includegraphics[angle=0,width=\linewidth]{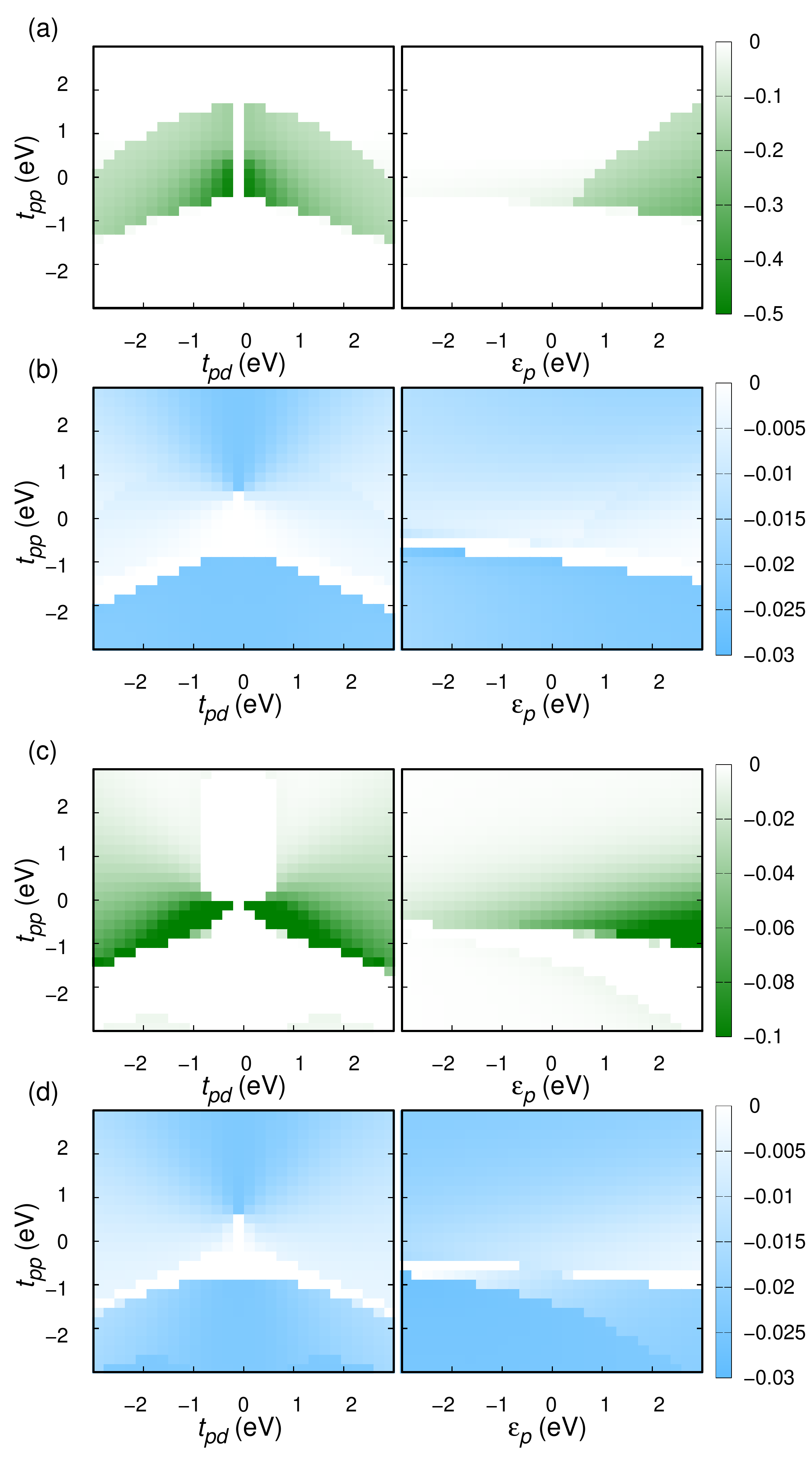}
\caption{Spin-spin correlations for zero doping with respect to $t_{pd}$ and $t_{pp}$ ($\epsilon_p$ = 2.2 eV) as well as $\epsilon_p$ and $t_{pp}$ ($t_{pd}$ = 1.3 eV). (a) nearest neighbour Cu-Cu correlation ($\langle S_i^dS_j^d \rangle$) (b) nearest neighbour O-O correlations ($ \langle S_i^pS_j^p \rangle$). (c) and (d) are the corresponding cases for doping $\delta$ = 0.25 (5 holes).}
\label{fig:ss}
\end{figure}

 Fig.~\ref{fig:ss} display the spin-spin correlations for ED of four-copper-sites cluster using $U_d$ = 8 eV, $U_p$ = 4 eV.  Note that here we neglect the small positive values for Cu-Cu correlation. 

\section{Magnon spectrum}
\label{app:mag}
\subsection{Model}
We study the effective Heisenberg-like spin-only model 
\begin{align}
H_\text{spin} &=  \sum_{\langle ij\rangle} J_{\alpha\beta} \vec{S}_{i\alpha } \vec{S}_{j\beta} \nonumber\\
&= \sum_{\langle ij\rangle} J_{\alpha\beta}\left(S^z_{i\alpha} S^z_{j\beta} + \frac{1}{2}S^+_{i\alpha}S^-_{j\beta} + \frac{1}{2}S^-_{i\alpha}S^+_{j\beta}\right)
\label{eq:H_heisenberg}
\end{align}
which emerges in the large $U$ limit from eq.~(1) of the main text. It includes AFM exchange $J_{dd}>0$ between the spin-$1/2$ located at the positions of the Cu atoms, AFM exchange $J_{pp}>0$ between the spin-$1/2$ located at the positions of the O atoms and a interaction $J_{pd}$ between O and Cu moments. $S^\nu_{i\alpha}$ are the spin-1/2 operators and $S^\pm_{i\alpha}$ the raising and lowering operators.

While we take the spin model as an effective phenomenological model we can estimate the spin exchange energy scales from perturbation theory: 

\begin{enumerate}
    \item A textbook result, e.g. see \cite{Lau2011}, gives for the positive charge transfer limit: $J_{pd} = 2 t_{pd} \Big( \frac{1}{U_p + \varepsilon_p} + \frac{1}{U} \Big)$.
    \item Similarly, in the positive charge transfer limit: $J_{pp} = \frac{4 t_{pp}^2}{U_p} $, {\it cf.} \cite{Kaushal2024}. Note that the higher order terms $\propto t_{pp} t^2_{pd}$ (i.e. going `on a O-Cu-O' triangle) can in principle be considered but they will: (i) be smaller for the positive charge transfer limit, and (ii) will depend on the occupancy of the `middle' copper sites (i.e. probably will vanish once this site is occupied, see below).
    \item Estimating $J_{dd}$ is more difficult. Once the oxygen is {\it not} occupied, we obtain the textbook result $J_{dd} = \frac{4t^4_{pd}}{\varepsilon_p^2}
    \Big(\frac{2}{U_p+ 2 \varepsilon_p} + \frac{1}{U}\Big)$ (valid in the positive charge transfer limit)~\cite{Zhang1988, Lau2011}. However, once the oxygen is occupied {\it and}
    assuming that the middle spin {\it does not} actively participate in the spin exchange~\footnote{This would lead to the three-spin interaction}, then there is {\it no} spin-dependent interaction (there merely is a lowering of the total energy due to the virtual processes -- however, these are spin independent).
    In fact, this result stays in line with what is written in \cite{Lau2011} [see Eq.~(5) in that paper].
\end{enumerate}

\subsection{Linear spin-wave theory}
We evaluate the magnon spectrum assuming the ordered groundstate of the $(\pi,\pi)$-AM, shown in Fig. 1 of the main text. We introduce 2 sublattices $\lambda$, employ a Holstein--Primakoff transformation, and take the large $S$ limit. If the spin at $i,\alpha,\lambda$ is $\uparrow$ the transformation reads
\begin{align}
    S^z_{i\alpha\lambda} &= S_\alpha-a^\dagger_{i\alpha\lambda} a_{i\alpha\lambda} \nonumber \\
    S^+_{i\alpha\lambda} &= \sqrt{2S_\alpha} a_{i\alpha\lambda} \nonumber \\
    S^-_{i\alpha\lambda} &= \sqrt{2S_\alpha} a^\dagger_{i\alpha\lambda} 
\end{align}
and
\begin{align}
    S^z_{i\alpha\lambda} &= -S_\alpha+a^\dagger_{i\alpha\lambda} a_{i\alpha\lambda} \nonumber \\
    S^+_{i\alpha\lambda} &= \sqrt{2S_\alpha} a^\dagger_{i\alpha\lambda} \nonumber \\
    S^-_{i\alpha\lambda} &= \sqrt{2S_\alpha} a_{i\alpha\lambda} 
\end{align}
for a $\downarrow$ spin. The operator $a_{i\alpha\lambda}$ ($a^\dagger_{i\alpha\lambda}$) annihilates (creates) a magnon, i.e. a spin flip, at $i,\alpha,\lambda$.

Keeping only terms linear or quadratic in $S_\alpha$ 
\begin{align}
H_\text{spin} =&  -4J_{dd} S_d^2 -8 J_{pp} S_p^2 + \nonumber \\
&+\sum_{\vec{k}} \vec{\eta}^\dag_{\vec{k}L} h_L(\vec{k}) \vec{\eta}_{\vec{k}L} + \vec{\eta}^\dag_{\vec{k}R} h_R(\vec{k}) \vec{\eta}_{\vec{k}R} + \mathcal{O}(S_\alpha^0)
\end{align}
which simplifies for $S_\alpha=1/2$ to 
\begin{align}
H_\text{spin} =&  -J_{dd} -2J_{pp} + \sum_{\vec{k}} \vec{\eta}^\dag_{\vec{k}L} h_L(\vec{k}) \vec{\eta}_{\vec{k}L} + \vec{\eta}^\dag_{\vec{k}R} h_R(\vec{k}) \vec{\eta}_{\vec{k}R}.
\label{eq:H_magnon_spectrum}
\end{align}
We defined $a^\dagger_{\vec{k}\alpha\lambda} = N^{-1/2} \sum_j \e^{\ii \vec{k} \vec{R}_{j\alpha\lambda}} a^\dagger_{j\alpha\lambda}$ where $\vec{R}_{j\alpha\lambda}$ is the position of a site $j,\alpha,\lambda$
and the fields
\begin{align}
\vec{\eta}_{\vec{k}R} &= (a_{\vec{k}dA}, a_{\vec{k}xA}, a_{\vec{k}xB}, a^\dag_{-\vec{k}dB}, a^\dag_{-\vec{k}yA},a^\dag_{-\vec{k}yB}) \\
\vec{\eta}_{\vec{k}L} &= (a_{\vec{k}dB}, a_{\vec{k}yA}, a_{\vec{k}yB}, a^\dag_{-\vec{k}dA}, a^\dag_{-\vec{k}xA},a^\dag_{-\vec{k}xB}).
\end{align}
The matrix elements are given by
\begin{align}
h_R(\vec{k}) =& \begin{pmatrix}
    h_{pd}(k_x)& h_{pd}(k_y) + h_0(\vec{k})\\
    h_{pd}^\dag(k_y) + h_0(\vec{k}) & h_{pd}(-k_y)
\end{pmatrix} 
\nonumber\\
&+ (J_{dd},J_{pp},J_{pp},J_{dd},J_{pp},J_{pp}) \1
\\
h_0(\vec{k}) =& 
\nonumber \\
& \hspace{-1cm}
\begin{pmatrix}
    \frac{J_{dd}}{2} (\cos k_x + \cos k_y) &0&0\\
    0&\frac{J_{pp}}{2}\cos \frac{k_x - k_y}{2}& \frac{J_{pp}}{2}\cos \frac{k_x + k_y}{2}\\
    0&\frac{J_{pp}}{2}\cos \frac{k_x + k_y}{2}& \frac{J_{pp}}{2}\cos \frac{k_x - k_y}{2}\\
\end{pmatrix} 
\\
h_{pd}(k) =& 
\frac{J_{pd}}{4}\begin{pmatrix}
    0&\e^{\ii k/2}&\e^{-\ii k/2}\\
    \e^{-\ii k/2}&0&0\\
    \e^{\ii k/2}&0&0\\
\end{pmatrix} 
\end{align}
and $h_L((k_x,k_y)) = h_R((-k_y,-k_x))$. Note however that $h_0((k_x,k_y)) = h_0((-k_y,-k_x))$ such that the matrices differ only in terms which are proportional to $J_{pd}$. 

We diagonalize \eqref{eq:H_magnon_spectrum} by a numerical Bogoliubov transformation, i.e. we multiply $h_{R,L}$ with the matrix $\1 (1,1,1,-1,-1,-1)$ and then diagonalize this non-hermitian matrix. The positive eigenvalues yield the magnon spectrum of \eqref{eq:H_heisenberg}.

\subsection{Results}

The resulting magnon spectrum, see Fig.~\ref{fig:magnon_spectrum}, features two Goldstone modes, a flat mode which can be associated with the ordering of the oxygen moments and a steep mode from the copper moments, because $J_{dd} > J_{pp}$. Additionally, an optical mode with oxygen orbital character appears due to the fact that the AFM ordering of the oxygen moments is intra unit cell. We observe large chirality splitting of the magnon modes because we expect $J_{pd}>J_{pp}$. We expect that most of the spectral weight is located on the magnon branches with Cu orbital character. However, quantitative studies of the dynamic spin structure factor of more microscopic model remains a formidable task for future. 

\section{Values of the orthogonalisation factors}
\label{sec:orthofactors}

The orthogonalisation factors $\mu, \nu, \chi $ are given in Table~\ref{tab:orthofactors}.
As the values of the orthogonalisation factors are strongly diminished with increased distance between 
the unit cells, only values for CuO$_4$  cells being up to third neighbor apart from each other are given in the table.

\begin{table}
    \centering
    \caption{Dependence of the orthogonalisation factors $\mu, \nu, \chi $ with distance between the CuO$_4$ cells $j$. Table adopted from Ref.~\cite{Jefferson1992}.    
    }    
\begin{tabular}{|c||c|c|c|}
\hline
j & $\mu$ & $\nu$ & $\chi$ \\
\hline
\hline
0 & 0.958 & 0.727 & 0 \\
\hline
1 & -0.14 & -0.273 & -0.133 \\
\hline
$\sqrt{2}$ & -0.02 &  0.122 & 0 \\
\hline
2 & -0.02 & -0.064 & 0.041 \\
\hline
\end{tabular}
    \label{tab:orthofactors}
\end{table}

\bibliography{biblio}

\end{document}